\documentclass[conference]{IEEEtran}
\IEEEoverridecommandlockouts

\def\BibTeX{{\rm B\kern-.05em{\sc i\kern-.025em b}\kern-.08em
    T\kern-.1667em\lower.7ex\hbox{E}\kern-.125emX}}

\usepackage{mathpazo}
\usepackage{times}
\usepackage{amsmath}
\usepackage{amsfonts}
\usepackage{latexsym}
\usepackage{amssymb}
\usepackage{mathabx}
\usepackage{float}
\usepackage{upref}
\usepackage{theorem}
\usepackage{graphicx}
\usepackage{psfrag}
\usepackage{cite}
\usepackage{cancel}

\usepackage{color}
\usepackage{tikz}
\usetikzlibrary{arrows.meta,bending}
\makeatletter
\newcommand\curvearrowed@[3]
  {%
    \begin{tikzpicture}
      \node[inner sep=.05ex](a){\kern-.25ex#3\kern-.05ex};
      \draw[arrows={-Latex[#1]},line width=#2]
        (a.north west) to[bend left=45] (a.north east);
    \end{tikzpicture}%
  }
\newcommand\curvearrowed[1]
  {%
    \relax
    \ifmmode
      \mathchoice
        {\curvearrowed@{}{.4pt}{$\displaystyle #1$}}
        {\curvearrowed@{}{.4pt}{$\textstyle #1$}}
        {\curvearrowed@{scale=.8}{.325pt}{$\scriptstyle #1$}}
        {\curvearrowed@{scale=.7}{.25pt}{$\scriptscriptstyle #1$}}%
    \else
      \curvearrowed@{}{.4pt}{#1}%
    \fi
  }
\makeatother

\usepackage{algorithm,algpseudocode}

\makeatletter
\newcommand{\removelatexerror}{\let\@latex@error\@gobble}
\makeatletter
\newcommand{\proofpart}[2]{%
	\par
	\addvspace{\medskipamount}%
	\noindent\emph{Part #1: #2}\par\nobreak
	\addvspace{\smallskipamount}%
	\@afterheading
}
\makeatother

\theoremstyle{plain}
\theorembodyfont{\normalfont\slshape}

\newtheorem{thm}{Theorem$\!$}
\newenvironment{theorem}
{\begin{thm}\hspace*{-1ex}{\bf.}}{\end{thm}}

\newtheorem{clm}[thm]{Claim$\!$}
\newenvironment{claim}{\begin{clm}\hspace*{-1ex}{\bf.}}{\end{clm}}

\newtheorem{lem}[thm]{Lemma$\!$}
\newenvironment{lemma}{\begin{lem}\hspace*{-1ex}{\bf.}}{\end{lem}}

\newtheorem{prop}[thm]{Proposition$\!$}

\newtheorem{cor}[thm]{Corollary$\!$}
\newenvironment{corollary}{\begin{cor}\hspace*{-1ex}{\bf.}}{\end{cor}}

\newtheorem{defn}[thm]{Definition$\!$}

\newtheorem{xmpl}[thm]{Example$\!$}
\newenvironment{example}{\begin{xmpl}\hspace*{-1ex}{\bf.}}{\hfill $\Box$ \end{xmpl}}

\newtheorem{cnstr}{Construction$\!$}

\newtheorem{rmk}[thm]{Remark$\!$}

\newcounter{enumrom}
\renewcommand{\theenumrom}{(\roman{enumrom})}

\makeatletter
\renewcommand{\@endtheorem}{\endtrivlist}
\makeatother

\makeatletter
\renewcommand{\thefigure}{{\@arabic\c@figure}}
\renewcommand{\fnum@figure}{{\bf Figure\,\thefigure}}
\makeatother

\newcommand{\cA}{\mathcal{A}}
\newcommand{\cB}{\mathcal{B}}

\newcommand{\cM}{\mathcal{M}}

\newcommand{\cO}{\mathcal{O}}

\newcommand{\cS}{\mathcal{S}}
\newcommand{\cT}{\mathcal{T}}

\newcommand{\ba}{\mathbf{a}}

\newcommand{\bb}{\mathbf{b}}

\newcommand{\br}{\mathbf{r}}
\newcommand{\bs}{\mathbf{s}}
\newcommand{\bt}{\mathbf{t}}
\newcommand{\bu}{\mathbf{u}}
\newcommand{\bv}{\mathbf{v}}

\renewcommand{\leq}{\leqslant}

\renewcommand{\geq}{\geqslant}

\newcommand{\Cref}[1]{Co\-ro\-lla\-ry\,\ref{#1}}

\outer\def\proclaim #1. #2\par{\medbreak
 \noindent{\bf#1.\enspace}{\sl#2\par}%
 \ifdim\lastskip<\medskipamount \removelastskip\penalty55\medskip\fi}

\begin{document}

\title{Reconstructing Mixtures of Coded Strings from Prefix and Suffix Compositions\\
\thanks{The work was funded by the DARPA Molecular Informatics, the NSF/SRC SemiSynBio program and the NSF grant CIF 2008125.}
}

\author{
    \IEEEauthorblockN{Ryan Gabrys, Srilakshmi Pattabiraman, Olgica Milenkovic}
\IEEEauthorblockA{ECE Department, University of Illinois, Urbana-Champaign, Urbana, IL, USA}
\IEEEauthorblockA{ryan.gabrys@gmail.com, \{sp16, milenkov\}@illinois.edu}
}

\maketitle

\begin{abstract}
The problem of string reconstruction from substring information has found many applications due to its relevance in DNA- and polymer-based data storage. One practically important and challenging paradigm requires reconstructing mixtures of strings based on the union of compositions of their prefixes and suffixes, generated by mass spectrometry readouts. We describe new coding methods that allow for unique joint reconstruction of subsets of strings selected from a code and provide matching upper and lower bounds on the asymptotic rate of the underlying codebooks. Under certain mild constraints on the problem parameters, one can show that the largest possible rate of a codebook that allows for all subcollections of $\leq h$ codestrings to be uniquely reconstructable from the prefix-suffix information equals $1/h$. 
\end{abstract}

\section{Introduction}

Modern digital data storage systems are facing fundamental storage density limits and to 
address the emerging needs for large data volume archiving, it is of great importance to identify new nanoscale recording media. Recently proposed molecular storage paradigms~\cite{al2017mass,goldman2013towards,grass2015robust,yazdi2017portable, yazdi2015rewritable,launay2020precise} offer storage densities that are order of magnitudes higher than those of flash and optical recorders but the systems often come with a prohibitively high cost and slow and error-prone read/write platforms. 
Polymer-based storage systems~\cite{al2017mass,launay2020precise} are particularly attractive due to their low cost and read latency~\cite{al2017mass}. In such platforms, two molecules that significantly differ in their masses represent the bits $0$ and $1$, respectively, and are used as building blocks in the process of synthesizing user-defined information content. The synthetic polymers are read by tandem mass (MS/MS) spectrometers. A mass spectrometer breaks multiple copies of the polymer uniformly at random, thereby creating prefixes and suffixes of the string of various lengths. The masses of these prefixes and suffixes are reported as the output of the system. If the masses of all prefixes from a single string are accounted for, reconstruction is straightforward. But if multiple strings are read simultaneously and the masses of prefixes and suffixes of the same length are confusable, the problem becomes significantly more complicated. It is currently not known which combinations of coded binary strings can be distinguished from each other based on prefix-sufix masses and for which code rates is it possible to perform unique multistring reconstruction.

In a related research direction, the problem of reconstructing a string from its MS/MS output was considered in~\cite{acharya2014string}, under the name of \textit{string reconstruction from its substring composition multiset}. The \textit{composition} of any binary string is the number of $0$s and the number of $1$s in the string. For example, the composition of $001$ equals $0^2 1^1$, indicating that $001$ contains two $0$s and one $1$, without revealing the order of the bits. The substring composition multiset, $C(\bs)$ of a string $\bs$ is the multiset of compositions of all possible substrings of the string $\bs$. As an illustration, the set of all substrings of $001$ is $\{ 0,0,1,00, 01, 001\}$, and the substring composition multiset of $001$ equals $\{ 0^1,0^1,1^1,0^2, 0^11^1, 0^21^1\}$. Two modeling assumptions are used in~\cite{acharya2014string} and subsequent works~\cite{pattabiraman2019reconstruction,gabrys2020mass, pattabiraman2020coding}: a) Using MS/MS measurements, one can uniquely infer the composition of a polymer substring from its mass; and b) When a polymer is broken down for mass spectrometry analysis, the masses of all its substrings are observed with identical frequency. 

Under the above modeling assumptions, the authors of~\cite{acharya2014string} established that strings are uniquely reconstructable up to reversal provided that the length of the strings $n$ is one less than a prime or twice a prime, or $n\leq 7$. The work~\cite{pattabiraman2019reconstruction,gabrys2020mass,pattabiraman2020coding} demonstrated that logarithmic code redundancy can ensure unique reconstruction of single strings drawn from codebooks based on Bertrand-Catalan paths or Reed-Solomon-like constructions.

However, the assumption that MS/MS output measurements contain the mass of all substrings is often not true in practice, as breaking the string in one rather than two position is easier to perform. In the former case, one is presented with masses of the prefixes and suffixes. Thus, for the string $001$, one would observe the multiset $\{ 0^1,\cancel{0^1},1^1,0^2, 0^11^1, 0^21^1\}$. Furthermore, in practice the content of multiple strings are often read simultaneously, which complicates the matter even further as it is not known a priori know which prefixes and suffixes are associated with a given string.

The problem addressed in this work may be stated as follows. We seek the size of the largest code $C(h)$ of binary strings of a fixed length with a \emph{$h$-unique reconstruction property} described next. For any subcollection $\bs_1, \bs_2, \ldots,\bs_{h'}$ of $h' \leq h$ strings from $C(h)$, one is presented with the union $\cM(\bs_1) \cup \cM(\bs_2) \cup \cdots \cup \cM(\bs_{\bar{h}})$ of the prefix-suffix composition multisets $\cM(\bs_i), \, i=1,\ldots,h',$ of the individual strings $\bs_i, \, i=1,\ldots,h$. The prefix-suffix composition multiset $\cM(\bs)$ of a string $\bs$ captures the weights of prefixes and suffixes of the string $\bs$ of all lengths. Unique reconstruction refers to the property of being able to distinguish all possible $h'$-unions and nonambiguously determine the identity of the strings in the collection. Our main result establishes that asymptotically and under some mild parameter constraints, the rate of the code $C(h)$ equals $1/h$. The proofs of our results and the code constructions rely on the use of binary $B_h$ strings, for which we establish additional results previously unreported in the literature.

The paper is organized as follows. Section~\ref{sec:intro} introduces the problem, the relevant terminology and notation. Section~\ref{sec:lbmc} describes the code constructions and the corresponding lower-bound analysis for the code rate. Upper bounds are presented in Section~\ref{sec:ubmc}. Due to space limitations, some of the proofs are relegated to the full version of the paper.

\section{Problem Statement and Preliminaries}\label{sec:intro}

We start by introducing the relevant notation. Let $\bs= s_1 \ldots s_n \in \{0,1\}^n$ be binary string of length $n$ and let  $\cM(\bs)$ denote the composition multiset of all prefixes and suffixes of $\bs$. For example, if $\bs = 01101$, then 
\begin{align*}
    \cM(\bs) = \Big \{ 0, 01, 01^2, 0^21^2, 0^21^3, 1, 01, 01^2, 01^3, 01^3 \Big \}.
\end{align*}
We denote the set of prefix and suffix compositions of $\bs$ as $\cM_p(\bs)$ and $\cM_s(\bs)$, respectively. For the above string, $\cM_p(\bs) = \{ 0, 01, 01^2, 0^21^2, 0^21^3 \}$ and $\cM_s(\bs) = \{ 1, 01, 01^2, 01^3, 01^3  \}$.

We seek to design a binary codebook $C(n,h) \subseteq \{0,1\}^n$ so that for any collection of distinct strings $\{\bs_1, \bs_2, \ldots, \bs_{\bar{h}} \} \in C(n,h)$ with ${\bar{h}} \leq h,$ the multiset 
\begin{align}\label{eq:constraint}
    \cM(\bs_1) \cup \cM(\bs_2) \cup \cdots \cup \cM(\bs_{\bar{h}})
\end{align}
uniquely determines the individual strings in the collection. We refer to a code that satisfies~(\ref{eq:constraint}) as an \textit{\textbf{$h$-multicomposition code},} or an \textit{\textbf{$h$-MC code}} for short. 
For simplicity of notation, we sometimes use $\cM(\cS)$ to the describe multi-composition set for $\cS = \{\bs_1, \bs_2, \ldots, \bs_h\}$. We also say that $C_p (n,h) \subseteq \{0,1\}^n$ is an $h$-prefix code if for any two distinct sets of size $\leq h$, say $\cS_1, \cS_2 \subseteq C_p$, 
$ \cM_p(\cS_1) \neq \cM_p(\cS_2).$ 

The next claim establishes a useful connection between our problem setup and one related to determining sums of binary strings. 
\begin{claim}\label{cl:sumsets} Given $\cM_p(\bs_1) \cup \cM_p(\bs_2) \cup \cdots \cup \cM_p(\bs_h)$, we can determine $\bs_1 + \bs_2 + \cdots + \bs_h.$ 

\end{claim}
\begin{IEEEproof} We prove the result for the case when $h=2$, as the generalization is straightforward. Suppose that $\bs_1, \bs_2 \in \{0,1\}^n$. Then, for given $\cM_p(\bs_1) \cup \cM_p(\bs_2)$, let $c_i$ denote the total number of ones in all compositions of length $i$ in the given multiset. It is straightforward to see that for $\bs_1 + \bs_2  = t_1 t_2 \ldots t_n$, one has $t_i = c_i - c_{i-1}$, with $c_0=0$.
\end{IEEEproof}

In light of the previous claim, it is straightforward to see that if we can recover $\cM_p(\bs_1) \cup \cM_p(\bs_1)$ from $\cM(\bs_1) \cup \cM(\bs_2)$ then we can uniquely determine the positions where $\bs_1$ and $\bs_2$ agree. The more challenging task is to determine the values of $\bs_1$ and $\bs_2$ at positions where $\bs_1$ and $\bs_2$ differ. The next example illustrates a setup where a pair of codestrings is confusable with another pair, and it highlights some of the challenges associated with the aforementioned problem.

\begin{example}\label{ex:notate} Suppose that $\bs_1= 0101101$ and $\bs_2 =  0110001$. Then

\begin{align*}
\cM_p &(\bs_1) \cup \cM_p(\bt_2) = \Big \{  0,0,01,01,0^21,01^2,0^21^2, \\
&  0^21^2, 0^21^3, 0^31^2, 0^31^3, 0^41^2, 0^31^4, 0^41^3 \Big \}. 
\end{align*}
Suppose we wish to recover $\{\bs_1, \bs_2\}$ from $\cM_p(\bs_1) \cup \cM_p(\bs_2)$. We can determine the first $4$ bits of both strings by inspection. At that point, we have that the strings $\{{ \bs_1,\bs_2\}}$ restricted to the first four positions equal 
$\{{0101,0110\}}.$

From $\cM_p(\bs_1) \cup \cM_p(\bs_2)$, we see that we need to either (i) append $0$ to $0101$ and $1$ to $0110$, or (ii) append $1$to $0101$ and $0$ to $0110$. Thus, we cannot tell if the pair of strings lies in $\{{0101001,0110101\}}$ or in $\{{0101101,0110001\}}.$

\end{example}

A binary $B_h$ sequence is a set $\cS_h(n)$ of binary strings of fixed length $n$ such that for any two distinct subsets of strings in $\cS_h(n)$, say $\cS = \{ \bs_1, \bs_2 , \ldots, \bs_{{\bar{h}}_1}\} \neq \cT=\{\bt_1, \bt_2, \ldots, \bt_{{\bar{h}}_2}\},$ where ${\bar{h}}_1, {\bar{h}}_2 \leq h$, one has
\begin{align}\label{eq:b2seq}
\sum_{\bs \in \cS} \bs \neq \sum_{\bt \in \cT} \bt.
\end{align}
Here, addition is performed over the reals. Thus, Claim~\ref{cl:sumsets} establishes two sufficient conditions for a set of strings to be an $h$-MC code: 
\begin{enumerate}
\item One can recover $\cM_p(\bs_1) \cup \cdots \cup \cM_p(\bs_h)$ from $\cM(\bs_1) \cup \cdots \cup \cM(\bs_h)$ for any h distinct codestrings $\bs_1, \ldots, \bs_h$; and 
\item The codestrings belong to a binary $B_h$ sequence $\cS_h(n)$. 
\end{enumerate}
These observations will be used to construct $h$-MC codes in Section~\ref{sec:lbmc}. Note that the condition that the codestrings in an MC code belong to a $B_h$-sequences is not necessary. For example, consider the case $\bs_1 = 011$, $\bs_2 = 000,$ $\bt_1 = 001$, $\bt_2 = 010$. Then,
$ \bs_1 + \bs_2 = 011 = \bt_1 + \bt_2,$
but $011 \in \cM(\bs_1) \cup \cM(\bs_2)$ and $011 \not \in \cM(\bt_1) \cup \cM(\bt_2)$, so that $\{\bs_1, \bs_2\}$ and $\{\bt_1, \bt_2\}$ are not confusable. However, a direct consequence of Claim~\ref{cl:sumsets} is that the maximum size of any $B_h$ sequence is at most the maximum size of a $h$-prefix code.

We show next that for sufficiently large code lengths, the maximum rate of an $h$-MC code is at least $\frac{1}{h}$. Furthermore, when $h$ is a power of two, we show in Section~\ref{sec:ubmc} that the maximum rate of an $h$-MC code is at most $\frac{1}{h}$. Note that from the earlier exposition, this implies that the maximum rate of a $B_h$ sequence is at most $\frac{1}{h}$, which improves upon the best currently known bound on $B_2$ sequence which is $.5753$~\cite{cohen2001binary}. For $h>2$, the authors are unaware of any known upper bounds on the size of binary $B_h$ sequences other the ones presented in this work.

\section{A Lower Bound on $h$-MC Codes}\label{sec:lbmc}

We start with a binary $B_h$ sequence and introduce redundancy into the underlying strings to ensure that given the multi-composition set of at most $h$ sequences, one can separate the prefixes from the suffixes. Then, given the set of prefixes, one can use the same idea in Claim~\ref{cl:sumsets} to recover the sum of the $h$ codestrings and hence the codestrings themselves.

Let $\cS_h(n) \subseteq \mathbb{F}_2^n$ be a $B_h$ sequence over $\mathbb{F}_2^n$. It is well-known that $\cS_h(n)$ can be constructed using the columns of a parity-check matrix of a code with minimum Hamming distance $\geq 2h+1$. Using this construction, we have that the rate of the resulting code for $n \to \infty$ satisfies
$$ \frac{1}{n} \log |\cS_h(n)| = \frac{1}{h}.$$

For our construction, we will also make use of the notion of a Dyck path: A string $\bs  \in \mathbb{F}_2^N$ is a Dyck path if $\text{wt}(\bs) = \frac{N}{2}$ and for $i \in [N-1]$,
\begin{align}\label{eq:dyck1}
\text{wt}(s_1 s_2\ldots s_i) \geq \Big \lfloor \frac{i}{2} \Big \rfloor + 1.
\end{align}

Our approach is to start with the set of strings $\cS_h(n)$ and generate the set $C(N,h) \subseteq \mathbb{F}_2^{N}$ from $\cS_h(n)$ so that the following two properties hold:
\begin{enumerate}
\item The set $C(N,h)$ is a $B_h$ sequence over $\mathbb{F}_2^N$, and
\item A string $\bs \in C(N,h)$ is a Dyck path. 
\end{enumerate}

The next claim establishes that if the code $C(N,h)$ satisfies these two properties, then it is an $h$-MC code.

\begin{claim}\label{cl:dyckrec} Suppose that $C(N,h)$ is a $B_h$ sequence where for any $\bs \in C(N,h)$, (\ref{eq:dyck1}) holds. Then, $C(N,h)$ is an $h$-MC code.
\end{claim}

To accomplish our goal, we perform a simple ``balancing procedure'' on each codesting $\bs \in \cS_h(n)$ and then append $\cO(\sqrt{n})$ bits of redundancy to the beginning and end of $\bs$ so that the resulting string has length $N = n + \cO(\sqrt{n})$. Note that under this setup, it follows that for any $\epsilon$, we have
$$ \frac{1}{N} \log |C(N,h)| = \frac{1}{n + \kappa \sqrt{n}} \log |\cS_h(n)| =  \frac{1}{h} - \epsilon,$$
where $\kappa$ is a constant, and $\epsilon>0$ can be made arbitrarily small for $n$ sufficiently large.

Rather than work directly with the weights of strings as described in (\ref{eq:dyck1}), we use the running digital sums (RDSs). For a string $\bs \in \mathbb{F}_2^n$, the RDS up to coordinate $i$ is defined as $R(\bs)_i = 2\text{wt}(s_1 s_2 \ldots s_i) - i$. If the subscript $i$ is omitted, then $R(\bs) = 2\text{wt}(\bs) - |\bs|$, where $|\bs|$ denotes the length of $\bs$. Note that using the running digital sum, (\ref{eq:dyck1}) can be rewritten as $\text{wt}(\bs) = \frac{N}{2}$ and $R(\bs)_i > 0,$ for any $i \in [N-1]$.

The balancing procedure operates as follows: Let $\bs \in \cS_h(n)$, and for simplicity, assume that $\sqrt{n}$ is an even integer. We begin by dividing up $\bs$ into blocks of length $\sqrt{n}$ so that $\bs = \bs_1 \bs_2 \ldots \bs_{\sqrt{n}}  \in \mathbb{F}_2^n,$ and $\bs_{j} \in \mathbb{F}_2^{\sqrt{n}}$. We introduce an auxiliary string $\bu=\bu_1 \ldots \bu_{\sqrt{n}} \in \mathbb{F}_2^n$ that is ``approximately'' balanced following an idea similar to Knuth's balancing. Initialize $\bu_1 = \bs_1$. For $\bs \in \mathbb{F}_2^m$, let $\overline{\bs} = \bs + 1\ldots 1 \in \mathbb{F}_2^m$ denote the result of flipping all the bits in $\bs$. Then, for $j \in \{2,3,\ldots,\sqrt{n}\}$, we define $\bu_j \in \mathbb{F}_2^{\sqrt{n}}$ as
\begin{align}\label{eq:procdbal}
\bu_j = \begin{cases}
\bs_j,& \text{ if } R(\bu_1 \ldots \bu_{j-1}) < 0, \text{ and } R(\bs_j) \geq 0,\\
\overline{\bs}_j,& \text{ if } R(\bu_1 \ldots \bu_{j-1}) < 0, \text{ and } R(\bs_j) < 0,\\
\bs_j,& \text{ if } R(\bu_1 \ldots \bu_{j-1}) \geq 0, \text{ and } R(\bs_j) < 0,\\
\overline{\bs}_j,& \text{ if } R(\bu_1 \ldots \bu_{j-1}) \geq 0, \text{ and } R(\bs_j) \geq 0.
\end{cases}
\end{align}

The first claim immediately follows from~(\ref{eq:procdbal}).

\begin{claim}\label{cl:bbal} For any $j \in [\sqrt{n}]$, $|R(\bu_1 \ldots \bu_j)| \leq \sqrt{n}.$
\end{claim}

We also have the following lemma.
\begin{lemma}\label{lem:rdsz} For any $i \in [n]$, $ |R(\bu)_i| \leq \frac{3}{2} \sqrt{n}.$
\end{lemma}

We now describe our encoder. Let $\bu \in \{0,1\}^n$ be the string which is the result of the procedure described in (\ref{eq:procdbal}), and suppose that $\br \in \{0,1\}^{\sqrt{n}}$ is such that for any $j \in [\sqrt{n}]$:
\begin{align}\label{eq:uzlabel}
\br_j = \begin{cases}
1, & \text{ if } \bu_j \neq \bs_j,\\
0, & \text{ if } \bu_j = \bs_j.
\end{cases}
\end{align}
From $\br$, we form a string $\bs \in C(N,h)$, where $N = n + \frac{17}{2} \sqrt{n} + 6$, and henceforth assume for simplicity that $N$ is an even integer. The following claim is used in our subsequent derivations.

\begin{claim}\label{cl:bal2} Let $\bv = {\bf1}^{5/2 \sqrt{n} + 1} \br \bu \in \{0,1\}^{n + 7/2 \sqrt{n} + 1}$. Then, for any $i \in [n + \frac{7}{2} \sqrt{n} + 1]$,
$$ |R(\bv)_i| \leq 5 \sqrt{n} + 1.$$
Furthermore, for any $i \in [n + 7/2 \sqrt{n}+1]$, 
$$ R(\bv)_i > 0. $$
\end{claim}

We now append redundant bits to the string $\bv$ described in Claim~\ref{cl:bal2} in order to get a string $\bs \in \{0,1\}^{N}$ which is a Dyck path. This results in the following claim.

\begin{claim}\label{cl:ydef} Let $N =n + \frac{17}{2} \sqrt{n} + 2$ be an even integer and let $\bv = {\bf1}^{5/2 \sqrt{n} + 1} \br \bu \in \{0,1\}^{n + 7/2 \sqrt{n} + 1}$ be as defined in Claim~\ref{cl:bal2}. Suppose that $w = \text{wt}(\bv)$. Then, the string 
$ \bs =  \bv {\bf1}^{\frac{N}{2} - w} {\bf0}^{\frac{N}{2}-(|\bv| - w)}  $
is a Dyck path.
\end{claim}

Now, assume that $C(N,h) \subseteq \mathbb{F}_2^N$ is constructed according to the procedure leading to Claim~\ref{cl:ydef} and once again assume that $N = n + \frac{17}{2} \sqrt{n} + 2$ is an even integer. The next theorem is the main result of this section and it establishes the correctness of our construction through the description of a simple decoding algorithm.

\begin{theorem}\label{thm:prf} Suppose that $\bs_1, \bs_2, \ldots, \bs_h \in C(N,h)$, where $C(N,h)$ is constructed according to the $B_h$ and balancing procedure. Then, given $ \cM(\bs_1) \cup \cM(\bs_2) \cup \cdots \cup \cM(\bs_h)$, we can uniquely determine $\{ \bs_1, \ldots, \bs_h \}$. Furthermore, for any $\epsilon >0$, there exists $n_\epsilon >0$ such that for all $N \geq n_\epsilon,$ $ \frac{1}{N} \log |C(N,h)| \geq  \frac{1}{h} - \epsilon.$
\end{theorem}
\begin{IEEEproof} For simplicity, we prove the result for $h=2$, but the extension for general values $h$ is straightforward. According to Claims~\ref{cl:dyckrec} and~\ref{cl:ydef}, we can recover $\cM_p(\bs_1) \cup \cM_p(\bs_2)$ from $ \cM(\bs_1) \cup \cM(\bs_2)$ since $\bs_1, \bs_2$ are Dyck paths. From $\cM_p(\bs_1) \cup \cM_p(\bs_2)$, we can recover $\bs_1 + \bs_2$ according to Claim~\ref{cl:sumsets}. Given that $\bs_1 =  {\bf1}^{5/2 \sqrt{n} + 1} \br_1 \bu_1 {\bf1}^{\frac{N}{2} - w_1}  {\bf0}^{\frac{N}{2} - (|\bv_1| - w_1)}$ and $\bs_2 =  {\bf1}^{5/2 \sqrt{n} + 1} \br_2 \bu_2  {\bf1}^{\frac{N}{2} - (|\bv_2| - w_2)}  {\bf0}^{\frac{N}{2} - z_2}$, from the first $n + \frac{7}{2} \sqrt{n} + 1$ coordinates of $\bs_1 + \bs_2$ we can recover
\begin{align*}
\left( \br_1 + \br_2, \bu_1 + \bu_2 \right) \bmod 2.
\end{align*}
Next, for shorthand, write $\bu = \bu_1 + \bu_2 \bmod 2 = \bu_1 \bu_2 \ldots \bu_{\sqrt{n}}$ and $\br = \br_1 + \br_2 \bmod 2 = r_1 \ldots  r_{\sqrt{n}}$. Define $\tilde{\bu} = \tilde{\bu}_1 \ldots \tilde{\bu}_{\sqrt{n}}$ so that for $j \in [\sqrt{n}]$, 
\begin{align*}
\tilde{\bu}_j = \begin{cases}
\bu_j, & \text{ if } r_j = 0,\\
\overline{\bu}_j & \text{ if } r_j = 1.
\end{cases}
\end{align*}
It is straightforward to verify from~(\ref{eq:procdbal}) that $\tilde{\bu} = \bs_1 + \bs_2 \bmod 2$. Since $\bs_1, \bs_2 \in \cS_2(n)$ are $B_2$ strings over $\mathbb{F}_2^n$, we can recover $\bs_1$ and $\bs_2$ from $\tilde{\bu}$, which concludes the proof.
\end{IEEEproof}

\section{Upper Bound on $h$-MC Codes}\label{sec:ubmc}

Next, we derive an upper bound on the maximum rate of an $h$-MC code. To this end, recall that $C_p \subseteq \{0,1\}^n$ is an $h$-prefix code if for any two subsets of sizes ${\bar{h}}\leq h$, say $\cS_1, \cS_2 \subseteq C_p$,  $ \cM_p(\cS_1) \neq \cM_p(\cS_2).$ 

Let $C^{(MC)}_h(n)$ be the largest possible $h$-MC code of length $n$ and suppose that $C^{(u)}_h(n)$ is the largest possible $h$-prefix code of length $n$. Formally, we use $R_h^{(MC)}$ to denote the capacity or the maximum asymptotic rate of an $h$-MC code, 
$$R_h^{(MC)} =  \lim_{n \to \infty} \sup \frac{1}{n} \log | C^{(MC)}_h(n) |.$$
We now show that when $h$ is a power of two, $R_h^{(MC)} \leq \frac{1}{h}$. Once again, for simplicity of exposition, we focus on the case $h=2$; a detailed treatment of the general is deferred to the full version of the manuscript.

The next lemma states that in order to derive an upper bound on the quantity $R_h^{(MC)}$, we can limit our attention to prefix codes. The result follows since the set of all suffixes is a function of the set of all prefixes provided the total number of ones in each codeword is the same and known beforehand. 

\begin{lemma}\label{lem:pset} For any $\epsilon >0$, there exists an $n_\epsilon > 0$ such that for all $n \geq n_\epsilon$, one has
$$\frac{1}{n} \log | C_h^{(MC)}(n) | \leq  \frac{1}{n} \log | C_h^{(u)}(n) | + \epsilon.$$
\end{lemma}

In what follows, we make use of a special type of $h$-prefix code, $C^{(r)}_{h,\ell}(n) \in \{0,1\}^n$, which we show in Lemma~\ref{lem:recur} has the same asymptotic rate as an $h$-prefix code. The code $C^{(r)}_{h,\ell}(n)$ may be viewed as a ``restricted'' $h$-prefix code with the special property that we can partition the codestrings of $C^{(r)}_{h,\ell}(n) \in \{0,1\}^n$ into $\ell$ blocks of equal length $\frac{n}{\ell}$ and each codestring $\bs \in C^{(r)}_{h,\ell}(n)$ can be written as
\begin{align}\label{eq:hcodea}
\bs = \bs_1 \bs_2 \bs_3 \ldots \bs_\ell,
\end{align} 
where $\bs_1, \ldots, \bs_\ell$ are codewords from a shorter, ``unrestricted'' $h$-prefix code of length $\frac{n}{\ell}$, denoted by  $C^{(u)}_{h}(\frac{n}{\ell}) \subseteq \{0,1\}^{\frac{n}{\ell}}$. The code $C_h^{(u)}(\frac{n}{\ell})$ is assumed to be a code of maximal rate. 
The following lemma establishes a connection between the rate of $C^{(r)}_{h,\ell}(n)$ and $C_h^{(u)}(n)$.

\begin{lemma}\label{lem:recur}
Let $C^{(r)}_{h,\ell}(n)$ denote the largest possible restricted $h$-prefix code of length $n$, and let $\ell=\cO(1)$. For any $\epsilon > 0$, there exists a $n_\epsilon > 0$, such that for all code lengths $n \geq n_\epsilon$,
$$ \frac{1}{n/\ell} \log |C_h^{(u)}(n/\ell)| \leq  \frac{1}{n} \log | C^{(r)}_{h,\ell}(n) | + \epsilon.$$
\end{lemma}

\begin{IEEEproof} Under this setup, strings in $C^{(r)}_{h,\ell}(n)$ are represented by a set of $\ell$ codewords from $C^{(u)}_h(\frac{n}{\ell}) \subseteq \{0,1\}^{\frac{n}{\ell}},$ where $C^{(u)}_h$ is an $h$-prefix code of length $\frac{n}{\ell}$ and maximal cardinality. In particular, the encoding for the code $C^{(r)}_{h,\ell}(n)$ works by representing our information as sets of $\ell$ codewords $\{ \bs_1, \ldots, \bs_\ell \} $ from the code $C_h^{(u)}(\frac{n}{\ell})$. Using this representation, there are $\binom{|C^{(u)}_h(n/\ell)|}{\ell}$ choices for a codestring in $C_{h,\ell}^{(r)}(n)$. Therefore, 
\begin{align*}
\frac{1}{n} \log | C^{(r)}_h(n) | = \frac{\log |\binom{|C^{(u)}_h(n/\ell)|}{\ell}| }{\frac{n}{\ell} \cdot \ell} &\geq \frac{\ell \log | C_h^{(u)}(n/\ell) |}{\ell \cdot \frac{n}{\ell}}  - \frac{\log \ell }{\frac{n}{\ell}} \\
&= \frac{\log | C_h^{(u)}(n/\ell)|}{n/\ell} - \epsilon
\end{align*}
provided that $n$ is large enough.
\end{IEEEproof}

Corollary~\ref{cor:capres} follows from Lemmas~\ref{lem:pset} and~\ref{lem:recur}.

\begin{corollary} The capacity of an $h$-MC code is at most the capacity of a restricted $h$-prefix code. \label{cor:capres}
\end{corollary}

As a consequence of the previous result, we can now turn our attention to deriving upper bounds on the capacity of restricted $h$-prefix codes $C_{h,\ell}^{(r)}(n) \subseteq \{0,1\}^n$. Again, we prove our claims for the case $h=2$ and defer the extension for $h > 2$ to the full manuscript.

\subsection{An Upper Bound on $2$-prefix Codes}

Our approach makes use of a partitioning idea similar to the one first described by Lindstrom~\cite{lindstrom1969determination} in the context of $B_2$ sequences. The main difference between our approach detailed and the one in~\cite{lindstrom1969determination} is that we make use of Lemmas~\ref{lem:pset} and \ref{lem:recur} to recursively improve our bound in the proof of Theorem~\ref{th:ub2}.

Once again, let $C^{(r)}_{2,\ell}(n)$ be a restricted $2$-prefix code of maximal size. For any $\bs \in C_{2,\ell}^{(r)}(n)$, we write $\bs$ as $\bs =  \ba \bb  \in C_{2,\ell}^{(r)}(n),$
where $\ba \in \{0,1\}^{a \frac{n}{\ell}}$ comprises the first $a$ blocks of $\bs$ and $\bb$ comprises the last $b = \ell - a$ blocks of $\bs$. Let 
\begin{align*}
B_{\ba} &= \Big \{ \bb \in \{0,1\}^{b \frac{n}{\ell}} :   \ba \bb \in C^{(r)}_{2,\ell}(n) \Big \},\\
 \cA &= \Big \{ \ba \in \{0,1\}^{a \frac{n}{\ell}} : \exists \bb \in \{0,1\}^{b \frac{n}{\ell}} \text{ s.t. } \ba \bb \in C^{(r)}_{2,\ell}(n) \Big \}.
 \end{align*}
The next two results will be used in the proof of Theorem~\ref{th:ub2}.

\begin{lemma}\label{lem:conf2} Suppose that $\ba, \ba' \in \cA$, $\ba \neq \ba'$, and that $\text{wt}(\ba) = \text{wt}(\ba')$. Then,
$$ \big |B_{\ba} \cap B_{\ba'} \big | \leq 1.$$ 
\end{lemma}
\begin{IEEEproof} Suppose, on the contrary, that $|\cB_{\ba} \cap \cB_{\ba'}| \geq 2$ and that $\cB_{\ba} \cap \cB_{\ba'} \supseteq \{ \bb_1, \bb_2\}$. In this case, we have
$$ \cM_p(\ba \bb_1) \cup \cM_p(\ba' \bb_2) = \cM_p(\ba \bb_2) \cup \cM_p(\ba' \bb_1).$$
To verify the above equality, note that all prefixes of length $a \frac{n}{\ell}$ have to be the same since $\cM_p(\ba \cup \ba') = \cM_p(\ba \cup \ba')$. Furthermore, since $\text{wt}(\ba) = \text{wt}(\ba')$, it follows that all prefixes of length greater than $a \frac{n}{\ell}$ have to be the same. 
\end{IEEEproof}

Next, let $\cA_w = \Big \{ \ba \in \cA : \text{wt}(\ba) = w \Big \}$ be the strings in $\cA$ that have weight $w \in [[a \frac{n}{\ell}+1]]$ (here, the notation $[[x]]$ is used for the set $\{{0,1,\ldots,x\}}$). 

\begin{claim}\label{cl:ubc2} For any $w \in [[ a \frac{n}{\ell}+1 ]]$,
$$\sum_{\ba, \ba' \in \cA_w}  \Big | B_{\ba} \cap B_{\ba'} \Big | \leq \big | C_2^{(u)}(n/\ell) \big|^{2a}. $$
\end{claim}
\begin{IEEEproof} By construction, each element of $\cA$ corresponds to the truncation of a codestring in $C_{2,t}^{(r)}(n)$ with respect to its first $a \frac{n}{t}$ coordinates. In particular, each element $\ba \in \cA$ can be written as:
\begin{align*}
\ba = \ba_1 \ba_2 \ldots \ba_a \in \{0,1\}^{a \frac{n}{\ell}},
\end{align*}
where $\ba_i \in C^{(u)}_{2}(n/\ell)$. Thus,
\begin{align*}
\sum_{\{\ba, \ba'\} \subseteq \cA_w} \Big | B_{\ba} \cap B_{\ba'} \Big | \leq \sum_{\{\ba, \ba'\} \subseteq \cA} 1 = \binom{|\cA|}{2} \leq \Big( |C_2^{(u)}(n/\ell) |^a \Big )^{2},
\end{align*}
as desired.
\end{IEEEproof}
We are now ready to prove our bound for $h=2$. 

\begin{theorem}\label{th:ub2}  For any $\epsilon > 0$, there exists an $n_\epsilon > 0$ such that for all $n \geq n_\epsilon$ one has $\frac{1}{n} \log |C_2^{(u)}(n)| \leq \frac{1}{2} + \epsilon$. 
\end{theorem}
\begin{IEEEproof} Recall that $\cA_w = \Big \{ \ba \in \cA : \text{wt}(\ba) = w \Big \}$
is the set of elements from $\cA$ that have weight $w \in [[a \frac{n}{\ell}+1]]$ and let 
$$ C^{(r)}_{w}(n) =  \Big \{ \ba \bb \in C^{(r)}_{2,\ell}(n) : \ba \in \cA_{w} \Big \}$$
denote the set of codewords from $C^{(r)}_{2,\ell}(n)$ that have $w$ ones in the first $a \frac{n}{\ell}$ coordinates. We first bound $|C^{(r)}_w(n)|$. It follows from the pigeonhole principle that there exists a $w^{*} \in [[a \frac{n}{\ell}+1]]$ such that
\begin{align*}
|C_{w^{*}}^{(r)}(n)| \geq \frac{1}{a \frac{n}{\ell} + 1} | C_{2,\ell}^{(r)}(n)|.
\end{align*}
Thus, $\frac{1}{n} \log |C^{(r)}_{2,\ell}(n)| \approx \frac{1}{n} \log | C_{w^{*}}^{(r)}|$ for $n$ sufficiently large.

Invoking Claim~\ref{cl:ubc2}, we have
\begin{align}
|C_{w^{*}}^{(r)}(n)| = \sum_{\ba \in \cA} | B_{\ba} | \leq& \Big | \cup_{\ba \in \cA} B_{\ba} \Big | + \sum_{\{\ba, \ba'\} \in \cA} | B_{\ba} \cap B_{\ba'} | \nonumber \\
\leq& 2^{b \frac{n}{\ell}} + \big | C_2^{(u)}(n/\ell) \big|^{2a}. \label{eq:r2mcb}
\end{align}
Using the straightforward bound $\big | C_2^{(u)}(n/\ell) \big|^{2a} \leq 2^{2 \frac{an}{\ell}}$ and setting $\frac{bn}{\ell} = \frac{2n}{3}$ so that $\frac{an}{\ell} = \frac{n}{3}$ shows that $\frac{1}{n} \log |C_{w^{*}}^{(r)}(n)| \leq \frac{2}{3}$ for $n$ large enough; as a result,  $\frac{1}{n} \log |C^{(r)}_{2,\ell}(n)| \leq \frac{2}{3}$, which implies $\frac{1}{m} \log | C_2^{(u)}(m) | \leq \frac{1}{m\ell} \log |C^{(r)}_{2,\ell}(m\ell)| + \epsilon \leq  \frac{2}{3} + \epsilon$ for any $\epsilon >0$ and $m$ large enough (see Lemma \ref{lem:recur}).

Next, we apply the same approach as above except that rather than using the weak bound $\big | C_2^{(u)}(n/\ell) \big|^{2a} \leq 2^{2 \frac{an}{\ell}}$, we now use the fact that $\frac{1}{m} \log | C_2^{(u)}(m) | \leq  \frac{2}{3}$ for $m$ large enough. This allows us to write $\big | C_2^{(u)}(n/\ell) \big|^{2a} \leq 2^{\frac{2an}{\ell} \cdot 2/3}$. In this case, based on (\ref{eq:r2mcb}), we set $\frac{bn}{\ell} = \frac{4}{7}$ and $\frac{an}{\ell} = \frac{3}{7}$ (so that $b \frac{n}{\ell}=2a\frac{n}{\ell} \cdot \frac{2}{3}$), which implies $\frac{1}{m} \log |C_2^{(u)}(m)| \leq \frac{4}{7} + \epsilon$ for any $\epsilon > 0$, provided that $m$ is large enough. Repeating this process, at each iteration we arrive at a smaller upper bound for $\frac{1}{m} \log | C_2^{(u)}(m)|$ from Lemma~\ref{lem:pset}. The proof follows by observing that $\frac{1}{m} \log | C_2^{(u)}(m)|= \frac{1}{2}$ is a fixed point of the iterative process.
\end{IEEEproof}

The next corollary follows from the previous theorem and Lemma~\ref{lem:pset}.

\begin{corollary} $R^{(MC)}_2 \leq \frac{1}{2}.$
\end{corollary}

\bibliography{IEEEabrv,biblio}
\bibliographystyle{IEEEtran}

\end{document}